\documentclass[runningheads]{llncs}
\usepackage{graphicx}
\usepackage{todonotes}
\usepackage[hidelinks]{hyperref}
\usepackage{amsmath}
\usepackage{amssymb}
\usepackage[caption=false]{subfig}
\usepackage{wrapfig}
\usepackage[numbers]{natbib}
\usepackage{tikz}
\usepackage[nameinlink]{cleveref}

\def\A{\mathcal{A}} 
\def\D{\mathcal{D}} 
\def\F{\mathcal{F}} 
\def\J{\mathcal{J}} 
\def\M{\mathcal{M}} 
\def\O{\mathcal{O}} 
\def\P{\mathcal{P}} 
\def\R{\mathcal{R}} 
\def\S{\mathcal{S}} 
\def\T{\mathcal{T}} 
\def\Z{\mathcal{Z}} 

\def\PI{{\boldsymbol{\pi}}}

\DeclareMathOperator*{\argmin}{arg\,min}

\newcommand*\circled[1]{\tikz[baseline=(char.base)]{\node[shape=circle,draw,inner sep=0.5pt] (char) {\textit{\textsf{#1}}}}}

\def\1{\protect\circled{1}}
\def\2{\protect\circled{2}}
\def\3{\protect\circled{3}}
\def\4{\protect\circled{4}}
\def\5{\protect\circled{5}}
\def\6{\protect\circled{6}}

\begin{document}
\title{Emergence in Multi-Agent Systems:\\ A Safety Perspective}

\author{
Philipp Altmann\inst{1}\and
Julian Sch\"onberger\inst{1}\and
Steffen Illium\inst{1}\and
Maximilian Zorn\inst{1}\and
Fabian Ritz\inst{1}\and
Tom Haider\inst{2}\and
Simon Burton\inst{3}\and
Thomas Gabor\inst{1}
}
\authorrunning{Altmann et al.}
\institute{
LMU Munich, Germany\and
Fraunhofer Institute for Cognitive Systems, Germany \and
University of York, United Kingdom\\
\email{philipp.altmann@ifi.lmu.de}}

\maketitle

\begin{abstract}
Emergent effects can arise in multi-agent systems (MAS) where execution is decentralized and reliant on local information. These effects may range from minor deviations in behavior to catastrophic system failures. To formally define these effects, we identify misalignments between the global inherent specification (the true specification) and its local approximation (such as the configuration of different reward components or observations). Using established safety terminology, we develop a framework to understand these emergent effects. To showcase the resulting implications, we use two broadly configurable exemplary gridworld scenarios, where insufficient specification leads to unintended behavior deviations when derived independently. Recognizing that a global adaptation might not always be feasible, we propose adjusting the underlying parameterizations to mitigate these issues, thereby improving the system's alignment and reducing the risk of emergent failures.
\end{abstract}

\section{Introduction}\label{src:introduction} 

Artificial intelligence (AI) is becoming prevalent in many practical applications, increasing the chance of multiple AI-based components interacting in the field, even when not designed or trained in conjunction.
We refer to such applications as \textit{AI Fusion}.
These AI components are treated similarly to extensive traditional software systems regarding their interactions.
However, as they are primarily applied in complex scenarios requiring an intelligent solution, this interaction might also cause unpredictable and surprising outcomes.
\footnote{In practice, AI components might be prone to solving simple tasks in exceedingly complex ways, which is directly associated with their perceived creativity, causing unconventional solutions, surprising to the user \cite{lehman2020surprising}.} 
Admittedly, to some extent, many of the surprises generated by AI originate from ill-defined goal functions that an AI adheres to more strictly than intended \cite{krakovna2020specification}.
In these cases, a human specification has a clear intent, and we would usually expect a human developer to follow this intent when a specification's formulaic description becomes unclear, thus counteracting incomplete specifications. 
From a software engineering perspective, employing AI entails living with oversight: 
Whenever we can fully specify the ideal behavior we want from a component, we can use that specification as a program without worrying about learning the intended behavior. 
Whenever we use AI, we want to look for solutions that we did not anticipate in their entirety. 
However, this also implies that AI might learn to exploit the blind spots in our specifications. 
\\[2pt]
Yet, while adapting to a validation task is feasible for aligning the local specification (i.e., the component's detached behavior), things become more elusive when combining several AI components into a global system. 
Behavior that does not violate our specification might still differ from another AI component's expectation (i.e., its learned priors). 
Thus, when two specification blind spots overlap, the regarded components might interact in new ways. 
This phenomenon where ``more is different''~\cite{Anderson72More} is generally referred to as \textit{emergence}. 
From a software standpoint, however, different might not always be preferred. 
Even though this global behavior might still fulfill its intended task, it might also cause minor defects or catastrophic failure.
\\[2pt]
In this work, we concentrate on multi-agent systems (MAS) due to their rigorous formalization to further investigate those effects in collective systems. 
As a running example, we use simple navigation tasks in two different gridworld environments that could be extrapolated to smart factories or large warehouse applications, where the safe and predictable operability of a collective system is inevitable for the successful application. 
Overall, we provide the following contributions: 
\begin{itemize}
  \item We introduce a formal model for tracing the emergent behavior to a misalignment between the intended global specification and its local approximation. 
  \item We provide two simple environments where such emergent behavior is prevalent in both planning and learning to find the optimal behavior. 
  \item We empirically show that by viewing the specified system as a surrogate of the intended target, its adaptation improves its alignment and helps mitigate such potentially catastrophic emergent effects to improve the system's safety. 
\end{itemize}

\section{Preliminaries}\label{sec:background}

\paragraph{\textbf{Emergence}}
Generally, emergence denotes a phenomenon where higher-level entities, patterns, and regularities arise through interactions among small or simpler entities that themselves do not exhibit such properties \cite{Anderson72More}.  
Honderich further defines emergent properties as unpredictable and irreducible so that an (emergent) property of a complex system arises out of the properties and relations characterizing its simpler constituents, while it is neither predictable from nor reducible to these lower-level characteristics \cite{honderich95oxford}.
Yet, understanding emergence has theoretical implications for how we model complex systems in science and engineering. 
We, therefore, aim to provide a formal model for tracing the origin of emergence in multi-agent systems.
Overall, models incorporating emergent properties can provide better management strategies for dealing with complex, adaptive systems, but they might also result in safety-critical deviation. 
According to Fromm~\cite{fromm2005types}, emergence can be further divided into four gradual types:
\begin{enumerate}
    \item[\textbf{I.}] \textbf{Simple/Nominal Emergence:} Comprises intentional and unintentional emergence with only \textit{feed-forward} relations and no top-down feedback.
    \item[\textbf{II.}] \textbf{Weak Emergence:} Includes simple positive or negative feedback.
    \item[\textbf{III.}] \textbf{Multiple Emergence:} Comprises more complex situations and adaptive systems with many kinds of feedback.
    \item[\textbf{IV.}] \textbf{Strong Emergence:} Arises in multiple levels from vast amounts of state variety (due to combinatorial explosion).
\end{enumerate}
Examples for \textit{type I} emergence include the intentional function of a software system emerging from its code or thermodynamic properties like pressure, volume, and temperature.
Also, allowing for top-down feedback, \textit{type II} might manifest in stable positive effects, like the locally coordinated foraging behavior of ants colonies or the self-organization of cells to form tissues. On the other hand, unstable effects, i.e., negative effects arising from positive feedback, might cause bubbles and crashes in stock and other financial markets.
Comprising comparatively simple effects, types I and II are said to be (in principle) predictable.
Type III, on the other hand, includes multiple kinds of feedback, such as emergent cooperation in social games like the prisoner's dilemma or adaptive systems like catastrophes in natural systems.
Type IV includes major evolutionary transitions, like the emergence of life or the emergence of culture.
Generally, this type occurs very rarely and at the boundaries of evolutionary systems. 
Within multi-agent systems, we mostly consider types II and III, where we focus on type III for the remainder of this work due to its non-predictable, multi-feedback nature.\\
Formally, we consider a system $\D$ comprised of $N$ autonomous agents $i$, where the behavior of each is described by its policy $\pi_i$, such that the collective behavior, or, joint policy $\PI$, can be represented by: 
\begin{equation}
\PI = \langle\pi_i\rangle^\otimes_{i\in\D} = \pi_1\otimes\dots\otimes\pi_{N},
\end{equation}
where $\otimes$ is the operator for combining components to systems (cf. \cite{wirsing2011ascens})\footnote{Note that instead of treating the environment as a different entity, we might as well model the environment as another agent in this formalism.}. Furthermore, we assume a specification $\F$, implemented as a boolean predicate, where $\F(\D) = \top$ iff all behaviors exhibited by the system $\D$ adhere to the specification $\F$, and, $\F(\D) = \bot$ otherwise. Note that specifications do not need to define full action sequences but might include very broad behavior ranges. Consequently, an \textit{emergent} property of system $\D$ is defined as:
\begin{equation}
\F(\PI) \neq \F(\pi_1)\land\dots\land\F(\pi_N)
\end{equation}
This implies that the collective behavior $\PI$ is not merely a sum or straightforward function of its elements but is the result of agent interactions. Consequently, the emergence property cannot be derived (or expected) solely from the properties of the individual policies $\pi_i$ but arises from their complex interplay.

\paragraph{\textbf{Multi-Agent Systems}}
To further formalize our scenario, we use a \textit{Markov decision process} $\langle \S, \mu, \A, \T, \R, \gamma \rangle$ \cite{puterman1990markov} with a set $\S$ of states $s_t$ at timestep $t$, with an initial state $s_0\sim\mu$, a set $\A$ of actions $a_t$, the transition probability $\T(s_{t+1}|s_t, a_t)$, a reward $r_t = \R(s_t,a_t)\in \mathbb{R}$, and a discount factor $\gamma$ to calculate the discounted return
\begin{equation}
G_t = \sum_{k=0}^{\infty} \gamma^k r_{t+k},
\end{equation}
where we consider the reward combining multiple objectives $\O$ (e.g., given by a previously defined specification), represented by boolean predicates, weighted by a vector $\P\in\mathbb{R}^{|\O|}$ such that the reward can be expressed as $\R=\P\O^T$ \cite{roijers2013survey}.
Furthermore, we consider a multi-agent system (MAS) formalized by a Markov game $\M=\langle\D, \S, \Z, \Omega, \J, \T, \R \rangle$, with a set $\D$ of $N$ agents $i$, a local observation function $\Omega:\S\to\Z^N$ to retrieve a tuple of local observations $\langle z_i \rangle_{i\in\D}$, a set $\J=\A^N$ of joint actions $\langle a_{t,i}\rangle_{i\in\D}$, and the joint reward function $\langle \R(s_t,a_{t,i})\rangle_{i\in\D}$.
The goal of each self-interested agent $i$ is to find an optimal strategy $\pi^*_i:\Z\to\A$ that maximizes the expected individual discounted return. From an agent's perspective, other agents are part of its environment, and policy updates by other agents affect the performance of an agent's own policy. 
The performance of policy $\pi_i$ of agent $i$ is estimated using a value function $V_i(s_t) = \mathbb{E}_\PI [G_{t,i} |s_t]$ for all $s_t\in\S$, based on the joint policy $\PI$ (cf. \cite{littman1994markov}).
Following our previous consideration of this collective behavior, we assume a system designer will define $\M$ according to the intended specification.

\paragraph{\textbf{Safety}}
The standard ISO 21448 defines the \textit{Safety Of The Intended Functionality} (SOTIF) as ``the absence of unreasonable risk due to a hazard caused by functional insufficiencies''~\cite{ISO21448:2022}.
These \textit{functional insufficiencies} can be caused by insufficiencies of the specification (e.g., the reward functions of the individual components) as well as performance insufficiencies (behavior deviations of each of the components from the intent) \cite{burton2023addressing}.
We consider this insufficient specification as a mismatch between the local specification and the global (intended) specification, which, as we argue, causes the diverging emergent behavior. 
In a collective system, which can be defined as an interconnected collection of components interacting in an environment to achieve a common target, this cause might also be noticeable as a performance insufficiency, where the learned behavior does not fulfill the local specification.
In any case, however, such emerging behavioral deviations violate the above prerequisites for SOTIF. 
Hendrycks et al.~\citep{hendrycks2021unsolved} explicitly consider the safety of machine learning systems and propose the unsolved problems of \textit{robustness}, \textit{monitoring}, \textit{systemic safety}, and \textit{alignment} regarding withstanding, identifying, and reducing hazards, as well as steering the fulfilled specification.
Consequently, we consider the misalignment of a system, caused by its insufficient specification, as the \textit{fault} causing an \textit{error} occurring in the form of a deviation from the intended global specification, ultimately exposing an emergent effect causing the \textit{failure} of the system.

\section{Emergence in Multi-Agent Systems}\label{sec:EMAS}
\begin{figure}[ht]\centering
\includegraphics[width=0.46\linewidth]{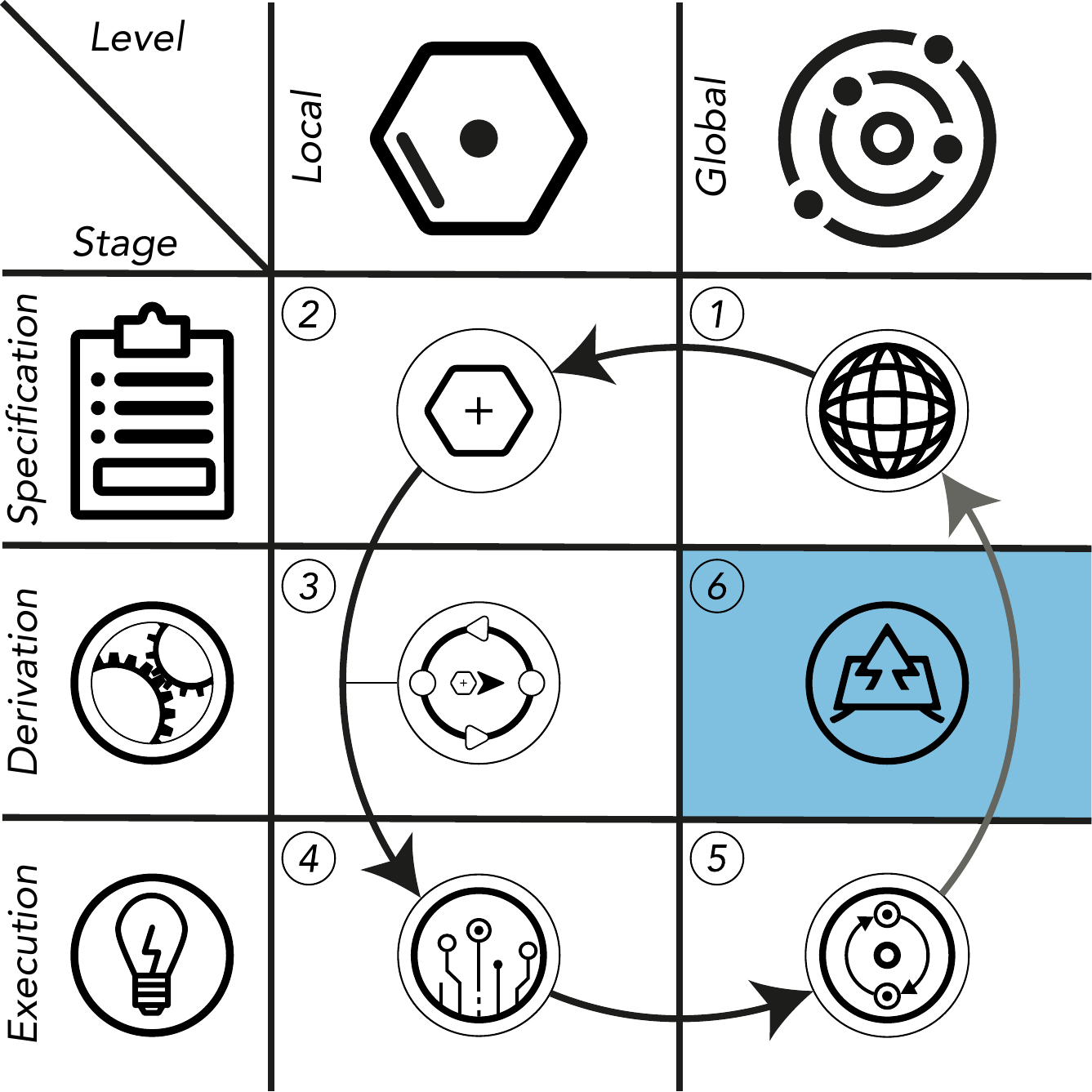}
\caption{\textbf{Emergence in MAS:} Generally, we assume a global task to be solved, given by the target specification $\F^*$ \1. For individual behavior (typically in the form of policies) to be derived \3, a local target (reward, observation, ...) needs to be formalized \2 (typically by defining $\M$). Optimally, the resulting policy \4 perfectly fulfills this previously defined local specification \2.  However, when executed in the global context \5, those policies might exhibit emergent effects \6. By adding this new step \6 to the MAS engineering cycle, we intend to discover emergent effects resulting from a misalignment between the \textit{real} target \1 and the resulting global behavior \5. We trace this effect to approximation errors induced throughout the overall development process, mainly to a miss-parameterization of the target to be optimized \2 diverging from the intended target inherent to \1.
}\label{fig:emas}
\end{figure}

We assume that a system's designer and/or user has a specific global behavior in mind.
We further assume that this desired behavior is accurately represented in the specification $\mathcal{F}^*$. 
From practical experience, however, we assume that the individual agent policy $\pi_i$ cannot be inferred directly from $\mathcal{F}^*$. 
One of the most prominent reasons is that agent interactions often have vastly complex effects, so we can only simulate the outcome of joint actions but not reverse-engineer the correct individual actions to reach a given joint goal. 
Furthermore, the technical setup often limits agents to accessing local information. Thus, they might not even have all the information required on a global scale to perform ideal actions according to the global specification $\mathcal{F}^*$.
\\[2pt]
We, therefore, assume that the designer defines the Markov game $\M$ accordingly, usually by choosing suitable observation and action spaces as well as reward parameterization $\P$.\footnote{We exemplify this process in \Cref{sec:Implementation} with a simple gridworld navigation task.}
Thus, we can only think of a global specification $\F^*$ as the result of a system specification $\hat{\F}$ that comprises entirely of individual agent specifications $\langle\hat{\F}_1,\dots,\hat{\F}_N,\rangle$. We acknowledge that although the individual specifications $\hat{\mathcal{F}}_i$ should be based on $\F^*$ by whoever defines them, they probably will not be able to capture an arbitrary complex global specification $\F^*$ entirely, but only approximate its behaviors to some extent. 
Formally, we can naturally define the global specification from the approximated individual specifications: 
\begin{equation}
\F^*\vdash\hat{\F}(\M|\pi_1\otimes\,\dots\,\otimes\pi_N) = \hat{\F}_1(\M|\pi_1)\,\land\, \dots \,\land\,  \hat{\F}_N(\M|\pi_N) 
\end{equation}
In most practical applications, this inaccuracy of local specifications is further exacerbated by the fact that agents usually do not perform fully up to specification. Instead, policy $\pi_i$ is also often derived from $\hat{\F}_i$ by a complex process (ranging from a human programmer team to a reinforcement learning artificial intelligence) and, while that process usually \textit{aims} to ensure that $\hat{\F}_i(\mathcal{M}|\pi_i) = \top$, we commonly end up with a policy $\pi_i$ that can only achieve $\hat{\F}_i(\mathcal{M}|\pi_i) \approx \top$, meaning that $i$ only fulfills $\hat{\F}_i$ ``most of the time'' or ``under certain conditions''.
\\[2pt]
Our theoretical argument shows that when we derive a joint policy $\PI = \pi_1 \otimes \dots \otimes \pi_N$ as it is commonly done\footnote{Note that in most settings, the joint policy returns a tuple of the actions returned by all individual policies. However, since in the general (partially observable) case we also need to adjust the observations passed on to each individual policy from the global state, we again use the $\otimes$ operator here and overload it to not only handle component composition but also state information decomposition and action composition, which are not inherently identical tasks but --- as we argue --- closely related tasks nonetheless.} from an originally global specification $\F^*$, we observe that we usually need to approximate twice: one time when splitting the global specification into local ones and another time when constructing local policies for the local specification. 
The latter of these approximations is a common process inherent to most software development or machine learning. Still, the former approximation is inherent to how multi-agent systems can currently be handled. 
This approximation gives rise to emergent effects: Even if we assume the original global concept to be perfect\footnote{Note that the perhaps even more common issue for developing any system is that we rarely have a perfect specification. We thus require an even earlier approximation at this step, i.e., we approximate the system we think we want via the specification we can actually write down. However, the inaccuracies of this approximation are again left to different subfields concerned about the whole variety of system design.}, we might lose the ability to perfectly represent it during the decomposition into local specifications, i.e., specifications at the single-agent level. 
Note that this \textit{multi-feedback} characteristic can also be connected to the nature of type III emergence introduced earlier. 
Specifically, the instances (i.e., the patterns in agent behavior) where the system $\M$ no longer fulfills the specification $\F^*$ while fulfilling the approximated joint specification $\hat{\F}$, which may happen because $\hat{\F}_1 \,\land\, \dots \,\land\, \hat{\F}_n = \hat{\F} \neq \F^*,$ are called \emph{emergence}. 
\\[0pt]
Fig.~\ref{fig:emas} summarizes the above considerations. 
Overall, we can consider $\M$ itself as an approximation of the intended specification $\F^*$.
Consequently, when deviating behavior emerges from the chosen parameterization of $\M$, its adaptation might be able to improve the decomposition of the local specifications such that the intended alignment $\hat{\F}\approx\F^*$ is corrected and emergent effects are mitigated.

\section{Implementation}\label{sec:Implementation}
\begin{figure}[ht]\centering
 \subfloat[\texttt{coin-quadrant}\label{fig:coin_quadrant}]{
    \includegraphics[width=0.3\textwidth]{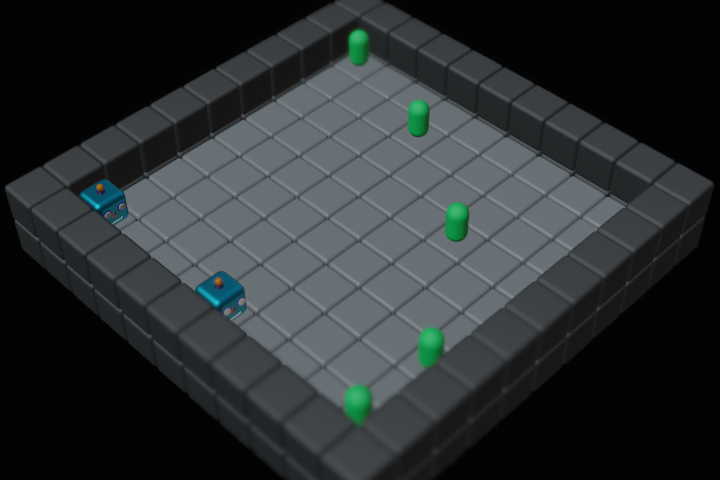}}
 \subfloat[Emergent Chasing\label{fig:chasing}]{
    \includegraphics[width=0.3\textwidth]{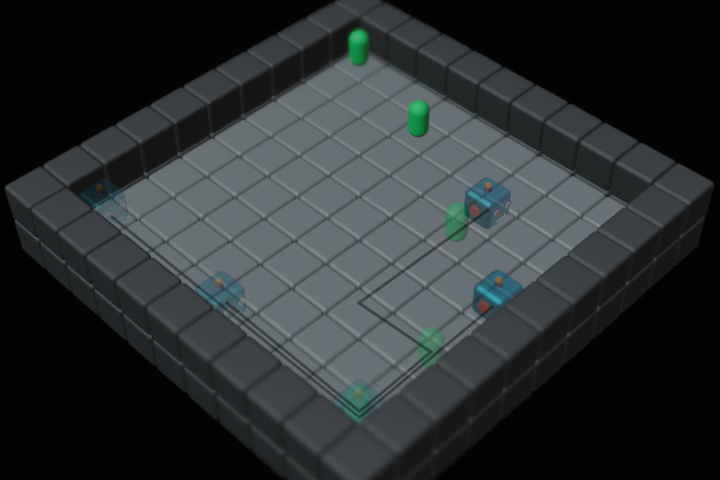}}
 \subfloat[Collective Behavior\label{fig:collective}]{
    \includegraphics[width=0.3\textwidth]{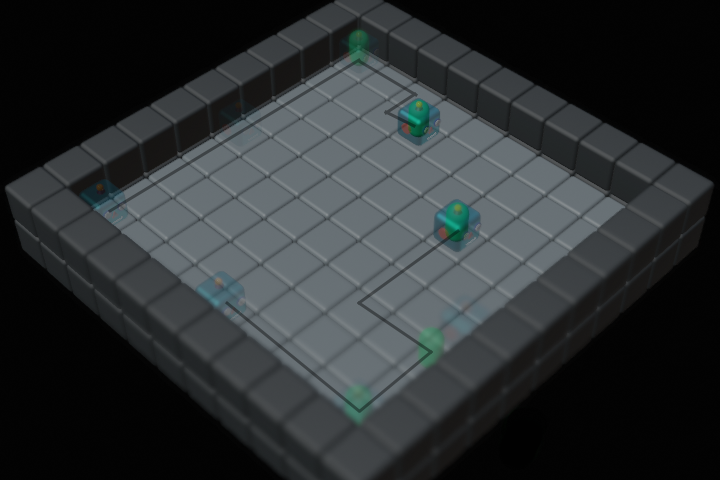}}\\
 \subfloat[\texttt{two-rooms}\label{fig:two_rooms}]{
    \includegraphics[width=0.3\textwidth]{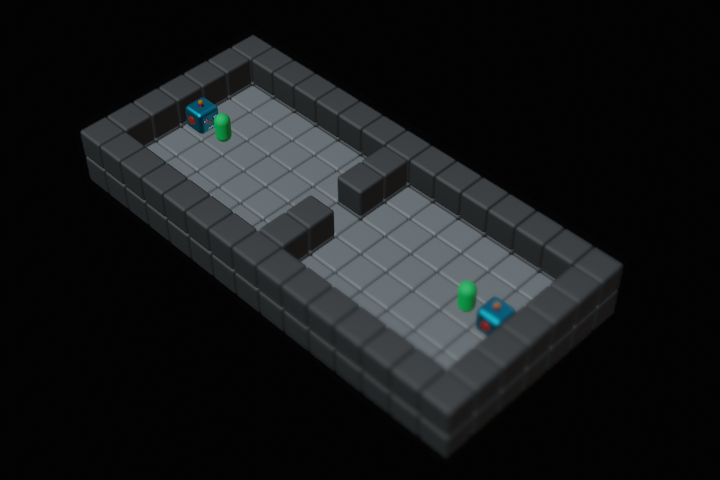}}
 \subfloat[Emergent Blocking\label{fig:blocking}]{
    \includegraphics[width=0.3\textwidth]{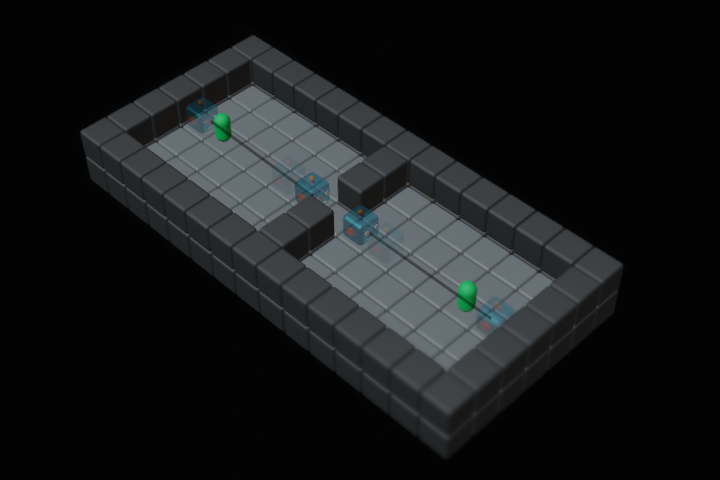}}
 \subfloat[Coordinated Behavior\label{fig:coordinated}]{
    \includegraphics[width=0.3\textwidth]{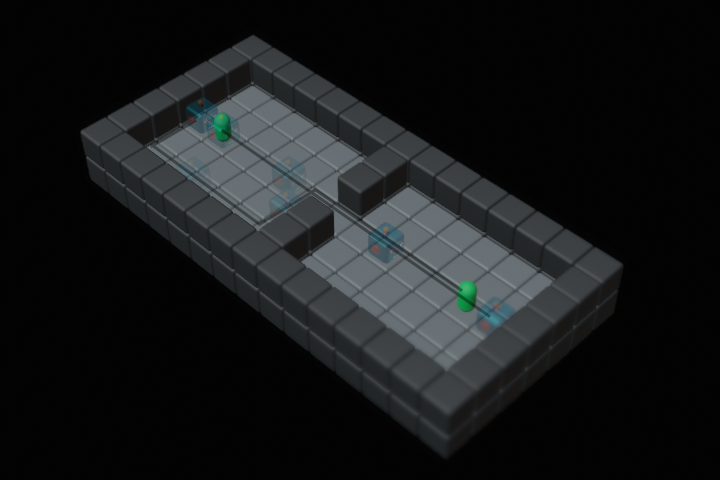}}
\caption{Overview of our emergence evaluation environments}\label{fig:environments}
\end{figure}

To illustrate our theoretical deliberations, we developed a framework explicitly for studying emergent phenomena in multi-agent systems based on simple gridworld environments to allow for in-depth visualization of the learned behavior\footnote{Refer to \url{https://github.com/philippaltmann/EMAS} for our full implementation.}. 
To provide a running example, we constructed two toy examples displayed in Figures~\ref{fig:coin_quadrant} and \ref{fig:two_rooms}. 
In both settings, two agents (blue) are tasked to fulfill a specific goal by collecting targets (green). 
Therefore, the global specification $\F^*$ is to collect their targets within as few steps as possible. 
In the left setting (\textit{coin-quadrant}), both agents need to cooperate to collect all targets. 
The second setting (\textit{two-rooms}) requires both agents to reach their opposing target by passing a bottleneck between the two agents' spawn points.
We abstract the two tasks into instances of a simple routing problem and use $\A=\{\uparrow, \rightarrow,\downarrow,\leftarrow\}$ encompassing four movement directions in state $\S$ defined as a $n\times m$ grid with $\text{cells}\in[\texttt{A1},\texttt{A2},\texttt{target},\texttt{field},\texttt{wall}]$.
Even though the tasks are seemingly simple independently, we aim to introduce complexity through the interaction dynamics between different entities and the entities themselves, with the specification reinforced via the chosen configuration.
To simulate insufficiencies of the chosen specification causing emergent behavior (i.e., $\F^*\neq\hat{\F}$), we allow extensive and precise specification of the task at hand.
While the state and action spaces remain comparably simple to ease learning and the overall evaluation, those specifications include the parameterization of a multi-objective reward signal and the definition of a suitable observation function. 
Similarly, \texttt{NetLogo} allows for simulating complex multi-agent phenomena \cite{wilensky2015introduction,tisue2004netlogo}. 
To provide compatibility with existing RL algorithms, our environments use the \texttt{gymnasium} API\cite{Towers_Gymnasium}.\vspace{-0.125cm}

\paragraph{\textbf{Parameterization}}
The reward consists of two objectives: (1) taking as few steps as possible to (2) reach the target state, implemented by the reward particles for taking any step, and reaching a target, parameterized by $\P=(-1,50)$, weighing the collection of targets higher than the step cost. 
Furthermore, we define the state $s_t$ to be observable by agent $i$ via 
$z_{t,i}=\Omega_i(s_t)$, with 
\begin{equation}\label{eq:observation}
\Omega_i(s_t) = \left\langle \rho(s_t,i), \argmin_{\texttt{target}\in s_t} | \rho(s_t,\texttt{target}) - \rho(s_t,i)| 
\right\rangle\in\mathbb{N}^4
\end{equation}
containing the current position of the concerned agent ($\rho(s_t, i)\in\mathbb{N}^2$) and the position of its immediate target\footnote{Note that, partial observability, besides being a common assumption in multi-agent reinforcement learning, has been shown to improve the agents' generalization to shifting environments \cite{Altmann23-CROP} and is commonly used for continuous robotic control tasks \cite{plappert2018multi}. Therefore, it could be considered a generally preferred implementation in practice.}, determined by their Manhattan distance ($|\cdot|)$.

\paragraph{\textbf{Agent types}}
We work with two types of agents. 
Planning agents implement a greedy \textit{traveling salesperson} (TSP) policy consisting of two steps. 
First, a complete graph is created based on the deterministic transitions $\T$ for all states $s\in\S$ and actions $a\in\A$, with the edge weights corresponding to the respective transition reward. 
Then, a greedy policy approximates the minimal Hamiltonian cycle on the constructed graph. Instead of traveling the whole cycle, execution is usually terminated when all coins are collected or all targets are reached. 
Note that while using the global state for deriving the policy, both \textit{TSP-agents} are derived locally, i.e., without observing the other agent. 
To represent a learning approach, we use \textit{reinforcement learning} (RL) agents based on the local observation $\O$.
Specifically, we use advantage actor-critic (A2C), a policy gradient approach \cite{mnih2016asynchronous}, where each policy and value approximation is represented by parameterized feed-forward neural networks updated using stochastic gradient accent on the expected return $\mathbb{E}_\pi(G_0)$. 
Overall, we train the agents independently and evaluate their interaction in conjunction, similar to an industrial application with agents from different vendors.

\paragraph{\textbf{Emergent Effects}}
The chosen environment configurations are the source of two unwanted emerging phenomena that lead to noticeable performance drops and potentially safety-critical system failures regardless of the agent type. 
In \textit{coin-quadrant}, the left agent is placed so that the costs of the shortest path to the upper left and lower right coin are equal. This will essentially lead to a random choice. Problems occur when this agent decides to go to the right coin first. Since the other agent is closer to the right coin, the agent will be the first to arrive and collect it. The same will happen for the second and the third coin, leading to a \textbf{chasing} behavior of the latter agent without it ever collecting any coin (cf. Fig.~\ref{fig:chasing}). This reduces the overall system performance since the resources spent by the latter agent are wasted. 
Worse than inefficient use of resources is the case in the \textit{two-rooms} environment, where the emergent behavior that the agents exhibit is both hindering performance and impeding the overall system functionality.
Due to the agents being placed at the same distance from the door, traveling the shortest path to their respective targets will result in a simultaneous arrival at this bottleneck. 
In our case, since both agents approach it at the same time, this leads to a \textbf{blocking} situation (cf. Fig.~\ref{fig:blocking}) where neither agent can pass the bottleneck, preventing them from ever reaching their designated target.

\paragraph{\textbf{Remedial Adaptations}}
Arguably, both effects generally emerge from the insufficient specification of ignoring the agents' interaction. 
However, given that deriving the behavior based on global information might not always be feasible, we aim to adapt our approximation of the target specification, defined by $\M$. 
In the following, we propose different approaches to address this misalignment for both planning and RL-agents that will be further evaluated in Section~\ref{sec:Evaluation}.
As the \textbf{TSP-agents}' behavior is solely based on the reward structure, we consider problem resolution via adaptation of the chosen reward parameterization ($\P:=(-1,50)$), applied to the reward particles ($\R_s=1$) for every step executed, and $\R_g=1$ for reaching the intended goal or $\R_g=0$ otherwise, where $\R=\P\cdot(\R_s,\R_g)^T.$
To integrate additional (potentially missing) information, we furthermore add the position of the agent (i.e., $x\in[1,n], y\in[1,m]$) into the parameterization $\P_c$ and scale the cost particle $\R_c$.
To prevent the emergence of the \textit{chasing} behavior in the \textit{coins-quadrant} environment, we suggest the following adapted parameterization, derived as a saddle point to resolve the cost equilibrium:
\begin{equation}
\P_{x,y} = (-0.55-0.05y, 50),
\end{equation}
introducing a linear gradient along the y-axis to the cost reward particle. 
Pushing the left agent towards the upper coin finally results in the intended collective behavior shown in Figure~\ref{fig:collective}.
Arguably, this linear gradient would also resolve other situations where an agent is equally distanced from two coins. 
However, preventing the emergence of the \textit{blocking} behavior turns out more complicated as the agents fail to solve the intended task. 
This failure could again be attributed to misalignment; in this case, missing information that not all steps might be attributed to equal costs, especially in bottleneck states.
To resolve this, we introduce the non-linear contortion of the cost parameterization along both axes 
\begin{equation}
\P_{x,y} = \left(\frac{5}{y}\sqrt[3]{\frac{x}{4}-1}-1,50\right),
\end{equation}
causing non-blocking optimal paths of different lengths, resulting in the coordinated behavior shown in Figure~\ref{fig:coordinated}.
While we will show these adaptations to be effective in preventing the emergent effects of TSP-agents, \textbf{RL-agents} require a different kind of approach. 
We argue that not only an insufficient reward function or cost structure can be the source of emergent behavior but also improper observations. 
As previously described, our observations contain information about the agent's location and current target. 
Even though the Manhattan-based target choice (cf. Eq.~\eqref{eq:observation}) approximately resembles the greedy policy of the TSP-agents, this target selection mechanism gives rise to the emergent chasing. 
Therefore, we propose to use an adapted distance metric $||\cdot||$ to prioritize targets $t$ whose path ($\mathbb{P}_i(t)$ according to \cite{bresenham1998algorithm}) for agent $i$ does not intersect any other agent $j\in\{\D\setminus i\}$:
\begin{equation}
||t - i|| = 
\begin{cases}
|\mathbb{P}_i(t)| + |\mathbb{P}_i(j)| \quad \text{if }\exists\,j\in\{\D\setminus i\} : j \in \mathbb{P}_i(t)\\
|\mathbb{P}_i(t)| \hspace{1.75cm} \text{otherwise}
\end{cases}
\end{equation}
Similarly, to prevent the \textbf{blocking} behavior, we propose adding an auxiliary target (e.g., at $(1,1)$) if the agents encounter an equal target distance to circumvent their collision. 
Note that reaching the auxiliary tile must be part of the training.

\section{Evaluation}\label{sec:Evaluation}
To verify that the presented adaptations for preventing emergent behavior indeed represent effective means for counteracting the experienced emergent behavior, we evaluate the different approaches in the following. 

\begin{figure}[!ht]\centering
 \subfloat[Chasing\label{fig:routes_chasing}]{ \includegraphics[width=0.25\textwidth]{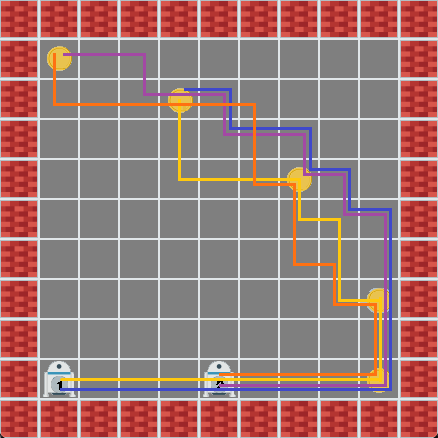}}
 \subfloat[Collection\label{fig:routes_collective}]{ \includegraphics[width=0.25\textwidth]{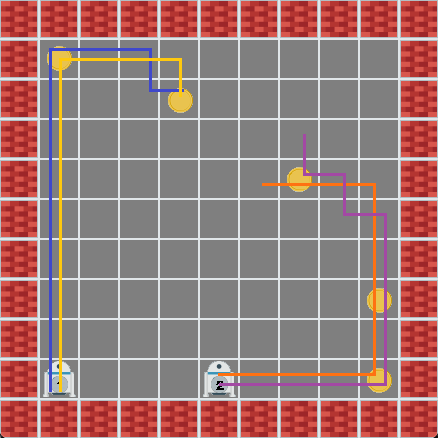}}
 \hspace{0.8cm}
 \subfloat[Collected Coins \\ \centering (randomized)\label{fig:collected_coins}]{ \includegraphics[width=0.36\textwidth]{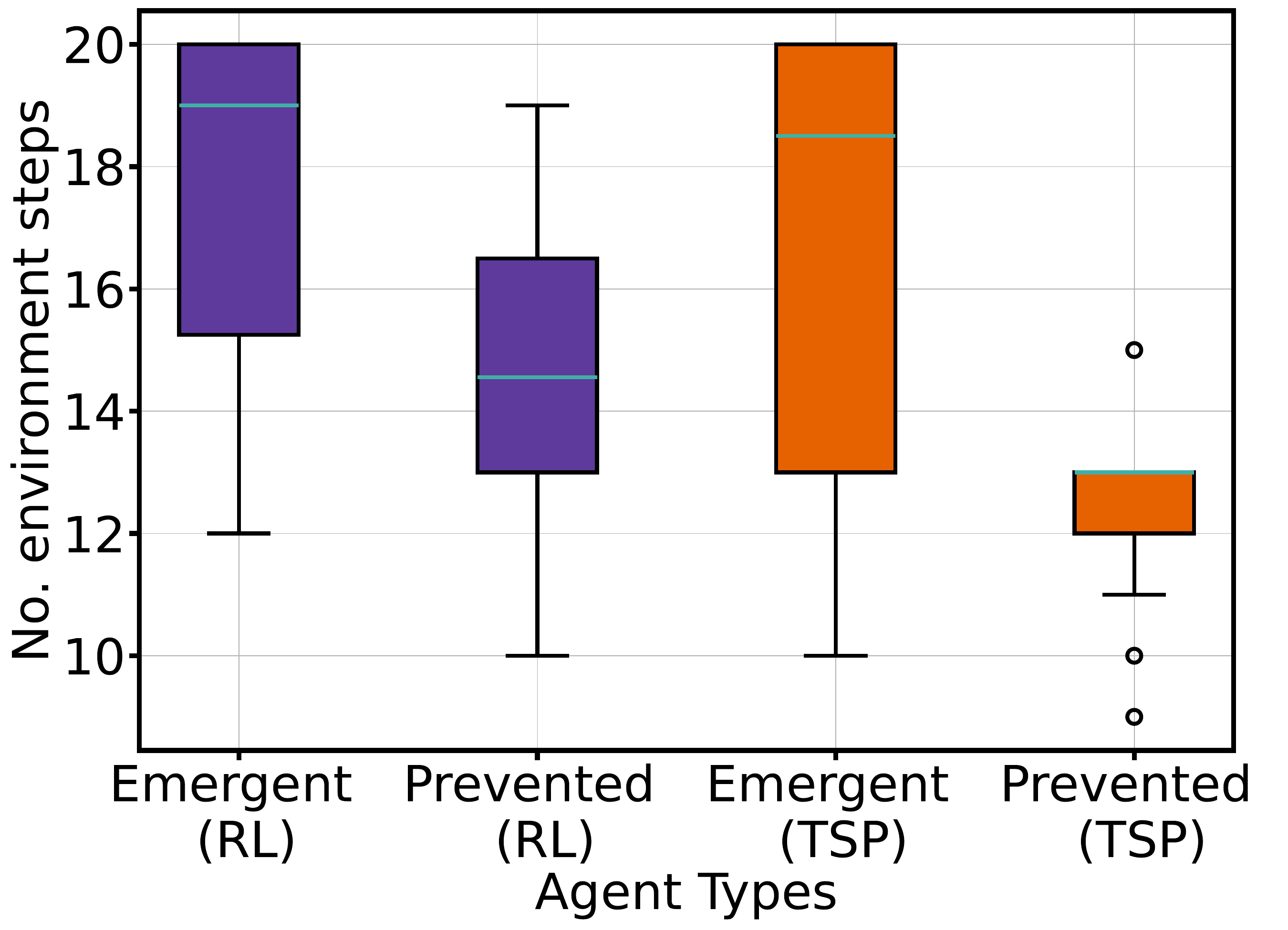}}\\
 \hspace{0.8cm}
 \begin{minipage}[c][0.45\textwidth]{.5\textwidth}\centering
 \subfloat[Blocking\label{fig:routes_blocking}]{\includegraphics[width=.7\textwidth]{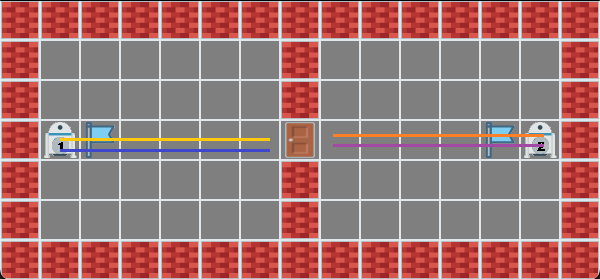}}\\\vspace{-0.3cm}
 \subfloat[Coordination\label{fig:routes_coordination}]{ \includegraphics[width=0.7\textwidth]{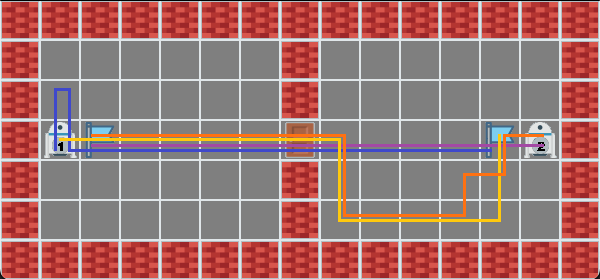}}
 \end{minipage}
 \begin{minipage}[c][0.4\textwidth]{0.35\textwidth}
 \subfloat[Reached Flags\label{fig:reached_flags}]{ \includegraphics[width=1.0\textwidth]{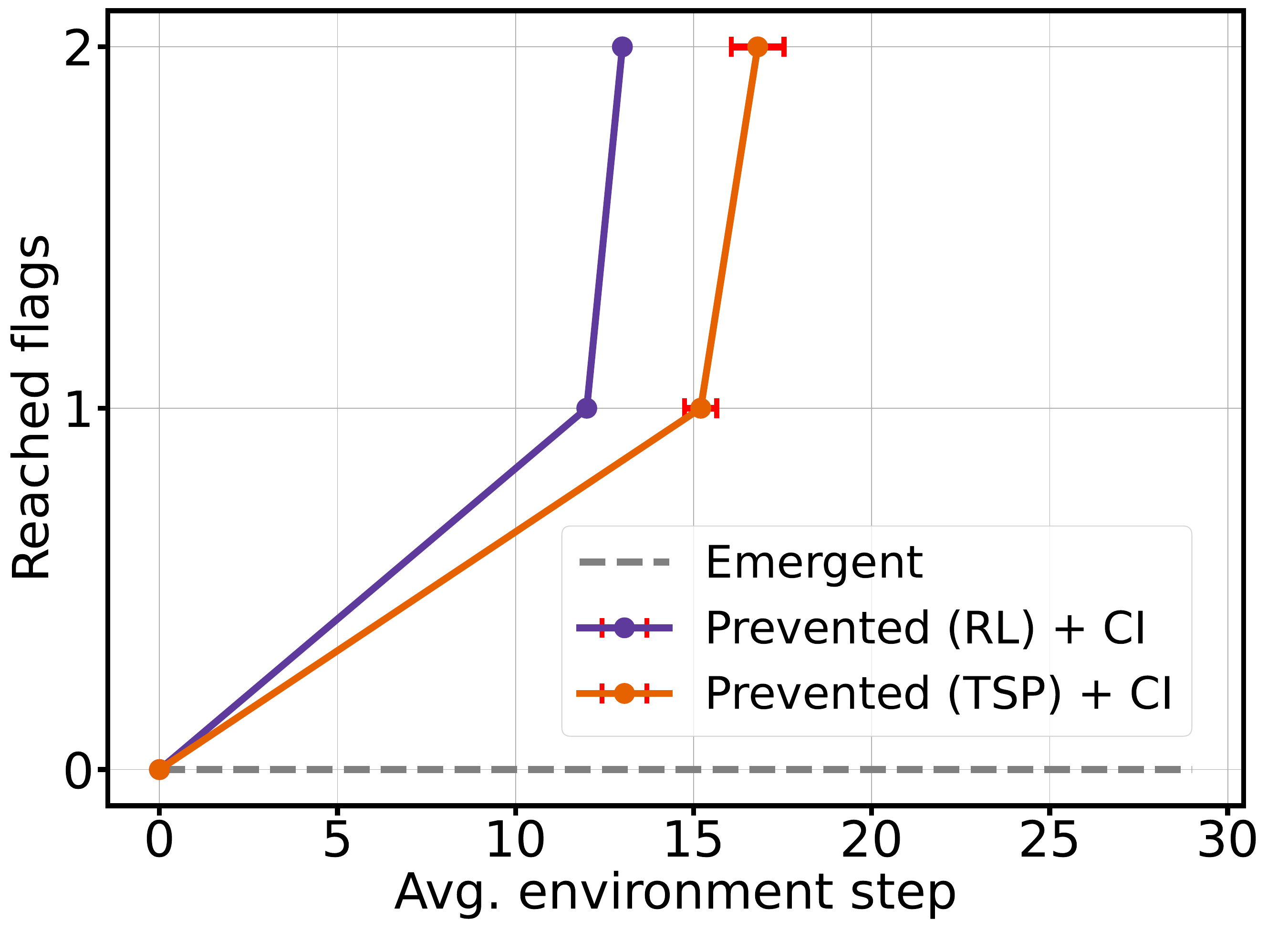}}
 \end{minipage}
\caption{
\protect\subref{fig:routes_chasing} and \protect\subref{fig:routes_blocking}  show one instance of the emergent behavior, whereas the routes in \protect\subref{fig:routes_collective} and \protect\subref{fig:routes_coordination} result from the usage of our emergence prevention approaches. 
The blue and purple trajectories correspond to the behavior of our RL-agents. 
The yellow and orange routes belong to the TSP-agents. \protect\subref{fig:collected_coins} and \protect\subref{fig:reached_flags} give an overview of the distribution of the required environment time steps to reach the targets considering multiple random seeds. For \textit{Collection}, we randomized the positions and the number of coins on the quarter circle. For \textit{Coordination}, we also plotted the confidence intervals (red whiskers).
}\label{fig:environments_behavior}
\end{figure}

\paragraph{\textbf{Collective Gathering}}
The emergent behavior described in Section~\ref{sec:Implementation} is clearly observable when considering the trajectories of the two agents in the \textit{coin-quadrant} environment. Figure \ref{fig:routes_chasing} shows the behavior of the different agent types for one example setting. Both agents aim for the bottom right coin first, but the left agent requires four additional steps to reach it. This time, Agent 2 is able to collect the initial coin and, following a zigzag-shaped route, collects one coin after another while Agent 1 purposelessly follows behind. Consequently, the total number of environmental steps required for collecting all coins is determined by the length of the route of Agent 2, which is 20 in this case. 
Figure \ref{fig:routes_collective} visualizes the change in behavior for Agent 1 when equipped with our emergence breaching mechanisms. As expected, this time, Agent 1 travels upwards instead. In the case of the RL-agent, that's because it detects Agent 2 on its path to the right coin and, therefore, adds +4 to the estimated path costs (the distance between Agent 1 to Agent 2), which changes the decision in favor of the top left coin. For the TSP-agent, the route change is caused by our adaptation of the underlying cost graph. Both solutions significantly affect the total number of required environmental steps for collecting all coins. Now, the work is distributed, with Agent 1 collecting the two top coins while Agent 2 takes care of the remaining three. 
To get an impression of the generalization abilities of the mechanisms to unseen coin arrangements, we generated 20 settings where 2--12 coins are randomly placed on the quarter circle. In 15 of these settings, both agent types showed emergent behavior. For the RL agent, we trained the model on the 5-coin setting (cf. Fig.~\ref{fig:routes_chasing}) using ten random seeds. We noticed that one of the seeds did not converge properly, so we did not consider it in the evaluation. Every trained model is then applied to the 20 generated coin settings, with and without the emergence prevention mechanism. Figure \ref{fig:collected_coins} shows the distribution of the required environment steps to gather all the coins across the different settings, averaged over the different policies. For both the RL and TSP-agents, we can see that usage of the prevention mechanisms reliably reduces the required step costs, thus effectively avoiding the additional expenses provoked by the emergent behavior, with a mean reduction of at least four steps. The TSP agent, on average, requires fewer steps than the RL agent, which already applies to the emergent case.

\paragraph{\textbf{Coordinated Navigation}}
If we keep the emergent phenomenon in \textit{two-rooms} untreated, the agents' trajectories end at the bottleneck (cf. Fig.~\ref{fig:routes_blocking}). The result is a deadlock, where no targets are collected, and the program is interrupted after the maximum number of environment steps (in our case, 30). 
The proposed solution for the RL-setting forces a detour of one of the agents before approaching the bottleneck. This guarantees that the lengths of the agent paths are different, shifting the point of contact away from the bottleneck. The left RL-agent gains two extra steps by traveling to its assigned auxiliary tile, followed by a turnaround, before approaching the target. The agent with the extended path determines the total environmental steps for the RL-agents. In our case, this corresponds to a mean of 13 steps (cf. Fig.~\ref{fig:reached_flags}). This exact behavior was observed repeatedly with different policies trained on ten random seeds.\footnote{Of the initial ten random seeds, the training only converged for 6, indicating the need to use more sophisticated training algorithms in the future.}
For the TSP-agents, the contortion of the cost parameterization leads to a lower cost path maneuvering around the center of the right room, extending the required number of steps for reaching the bottleneck. This adaptation gives the left agent enough time to reach the other room, thus resolving the deadlock situation as presented in Figure \ref{fig:routes_coordination}. For the TSP-agents, the contortion affects both agents equally, resulting in a mean of 16.8 steps to reach all flags (cf. Fig.~\ref{fig:reached_flags}). We observed a slight variance for the TSP-agents when approaching the second flag, with a 95\% confidence interval of $\approx (16.06, 17.54)$ steps. 

\paragraph{\textbf{Discussion}}
Note that for both environments, the presented trajectories for the TSP-agents are randomly selected from the set of routes with equivalent path costs. In contrast, the RL-agents routes are deterministically predetermined by choosing the respective action with the highest probability. That is, for example, why the route for agent 2 in the \textit{coin-quadrant} environment changes for the TSP-agent while it remains the same for the RL-agent. 
Considering the differences in performance of the agent types across the two emergent phenomena, we argue that the choice of the most suitable agent type and corresponding emergence prevention mechanism is likely task-dependent. Also, the mechanisms are not free of limitations. In the \textit{coin-quadrant} environment, for example, both agents might travel a path to the same coin, which intersects only immediately before the target. Thus, the extra costs are considered very late, which could potentially result in an emergent situation once again. Yet, on average, all our approaches proved to substantially reduce the number of required environment steps to reach the targets compared to the scenarios where the emergence is left untreated.
\\[2pt]
Overall, our two toy examples showed how emergence can manifest in multi-agent systems and how changes to its specification $\M$ can circumvent these effects. For instance, in the case of the \textit{coin-quadrant} environment, the target was to collect all coins in as few environment steps as possible. The initial local specification $\hat{\F}_{i}$ for the TSP-agents specified a uniform cost for all transitions, neglecting the positioning of the two agents and thus opening up the door for emergent behavior to occur. Based on this observation, we could reiterate over the MAS engineering cycle, acknowledging the fact that our current surrogate for the global specification appeared to be misaligned with our target, and introduce a gradient on the cost-function to account for the undesired agent interactions. The renewed execution did confirm that this adaptation was, in fact, successful in preventing the emergence.
In the following, we survey related approaches suitable for aligning the identified specification insufficiencies.

\section{Related Work}\label{sec:relatedwork}

\paragraph{\textbf{Alignment}}
Ensuring that autonomous agents follow the rules of their environment is crucial.
These rules are often part of the non-functional requirements, which, along with functional requirements, outline how a system should behave.
The field of alignment deals with the challenges of ensuring that learning systems adhere to these rules.
At a high level, \textit{AI Value Alignment} is concerned with aligning advanced AI systems, such as sophisticated autonomous agents, with human values~\cite{RUS19}.
Gabriel breaks down the challenges into two main types: technical and normative~\cite{G20}.
Technical challenges include figuring out how to teach agents values or principles so they consistently meet their goals.
A specific issue for more \textit{advanced} agents is \textit{reward hacking}, where agents find unexpected and sometimes unwanted ways to reach their goals~\cite{amodei2016concrete,LMK+17,LKE+18}.
Here, \textit{advanced} refers to the agents outperforming humans at solving specific tasks.
Normative challenges are about deciding which values or principles to teach agents.
Gabriel differentiates between \textit{minimalist} approaches, which aim to avoid uncertain outcomes by tying agents to a reasonable set of human values, and \textit{maximalist} approaches, which strive to align agents with the best human values on a broader societal or global level.
We concentrate on the technical challenges and use the term alignment in this narrower sense.
The work of Hendrycks et al.~\cite{hendrycks2021unsolved}, which we already took on in Chapter~\ref{sec:background}, refers to alignment as one of the major unresolved challenges in applying machine learning methods, where the difficulty lies in representing and optimizing complex human values. We agree with this.

\paragraph{\textbf{AI Safety}}
The technical side of alignment we focus on has also been explored in the context of \textit{AI Safety}, which focuses on preventing accidents and negative impacts from using advanced autonomous systems.
Amodei et al. outline five key problems in this area~\cite{amodei2016concrete}.
While they don't use the term alignment, later works~\cite{G20} and practical applications often cite these problems, especially \textit{negative side-effects} and \textit{reward hacking}~\cite{amodei2016concrete,LMK+17,LKE+18}, as examples of poor alignment.
Leike et al. introduce scenarios to test the safety features of agents, covering both the issues identified by Amodei et al. and additional agent-specific areas like exploration~\cite{LMK+17}.
They categorize problems into those of \textit{robustness} and \textit{specification}: Robustness issues arise when agents can't achieve their goals despite knowing what they are, while specification issues occur when agents don't fully understand their goals, often because they can't be fully captured through rewards.
A typical method to address this involves refining the model's outputs using human feedback after the initial training~\cite{OWJ+22}.
For example, people might rate the model's responses, creating a preference-based reward model for further training (see reinforcement learning from human feedback (RLHF)~\cite{CLB+17}).
Lately, RLHF has been most prominently used to align potentially unhelpful, toxic, or incorrect large language model responses with human expectations. 
Similarly, we could imagine using RLHF to identify and correct the local specifications' ``blind spots'' that lead to unwanted emergent behavior.
Overall, we interpret the problems identified by Amodei et al.~\cite{amodei2016concrete} as root causes of poor alignment.
Another line of work deals with detecting anomalies in RL systems \cite{sedlmeier2019uncertainty, haider2023out, haider2024can}. The goal is to recognize situations the agent has not encountered during training, which potentially lead to safety violations or other failure modes at runtime. In our example, anomaly detection could help uncover the fact that other agents are present during execution, which were not modeled during training, indicating underspecification of the system.

\paragraph{\textbf{Reward shaping}}
In sequential decision-making problems, goals are typically represented by a reward function, where the mathematical objective is to maximize the cumulative expected reward \cite{sutton2018reinforcement}.
Consequently, it makes sense to steer the learning of RL-agents by adjusting the reward function.
This is particularly relevant in problems where feedback is infrequent, for example, because the goal is initially difficult to achieve.
In multi-agent systems, reward shaping can also address the credit assignment problem, which denotes the estimation of the individual agents' contributions to the global reward.
For instance, Salimbeni et al. derive individual rewards based on \textit{Kalman filtering}~\cite{SMM+22}, and Wolpert et al. use the difference between the actual reward and an alternative reward that would be given if the average action of all agents were chosen~\cite{WT01}.
The alternative reward indicates whether the action chosen by the agent is beneficial or detrimental to the overall system.
Foerster et al.~\cite{FFA+18} further develop this approach within a centralized training with decentralized execution (CTDE) architecture.
They calculate the alternative reward by excluding the action of the agent under consideration and keeping the actions of the other agents unchanged, allowing for a more precise measurement of individual contributions.
The approaches above significantly change the rewards to encourage cooperation but may add unintended side effects.
An approach that neither changes the optimal strategy nor causes side effects that could enable reward hacking in a single-agent RL system is the potential-based reward shaping (PBRS) developed by Ng et al.~\cite{NHR99}.
PBRS supplements the reward at each timestep with the difference in potential between a state and its successor state.
The potential of individual states is usually defined in a separate function and contains expert knowledge about the problem to be solved, such as approximately optimal heuristics.
Similarly, we propose adapting the reward signal based on expert knowledge about the origin of the emergent effect when access to global information cannot be guaranteed.

\paragraph{\textbf{Multi-Objective Optimization}}
Functional and non-functional requirements naturally induce a multi-objective (MO) problem, thus increasing the complexity of specification-aware machine learning \cite{hayes2022practical, hayes2023brief}. As MO learning spans a Pareto-front of rewards for individual policies to optimize, `mixed' reward behaviors, i.e., locally Pareto-optimal strategies, quickly compound the potential emergence in multi-agent systems. In a cooperative setting with homogeneous agents, training with PBRS was shown to achieve compliance with functional and non-functional requirements, ultimately leading to proactively safe runtime behavior~\cite{RPM+22}. However, the majority of MO seems to require increasingly specific approaches such as Pareto Q-networks~\cite{van2014multi} or Pareto-conditioned networks~\cite{reymond2022pareto}. Since RL has only recently begun to cover MO problems (cf. \citet{felten_toolkit_2023}), we aim to further the field with our specification-based environment and evaluation.

\section{Conclusion}\label{sec:Conclusion} 

In summary, we formally addressed the critical issue of emergent behavior in multi-agent systems (MAS) by introducing a process model to trace this deviation from the intended specification to a misalignment induced throughout the approximative process. If not appropriately handled, those effects emerging from the conjunct execution of independent models can cause severe behavior and performance deviations and might even cause catastrophic failure. To illustrate our formal approach, we introduced two simple gridworld examples where insufficient specification causes emergent functional deviation and even failure. We argue that this emergence in MAS is induced via approximation errors throughout the modeling and learning process. A straightforward solution could, therefore, be the inclusion of missing information to improve said alignment, e.g., via human feedback. However, as global information might not always be a viable option in the applications of MAS, we propose alternative adaptations to the system model (i.e., our approximation) to mitigate those behavioral deviations.
This creates an iterative process where the definition of the system itself is adapted to (safely) conform to the intended global specification.
In our exemplary tasks, these adaptations include the contortion of the reward signal and the local observation. 
Experimental results validate our theoretical elaborations and the effectiveness of those adaptations. 
\\[1pt]
However, despite these promising results, the exemplary environments used in our experiments are relatively simple and may not capture the full complexity of real-world MAS applications. 
Additionally, the assumption that global specifications are accurately known and can be effectively approximated at the local level might not always hold in practical scenarios. 
Therefore, we encourage future work to focus on the iterative process of integrating potentially incomplete or ambiguous human specifications into the design of the system by connecting our methodology to human-in-the-loop approaches like RLHF. 
Also, developing more sophisticated benchmarks that connect our formal considerations to practical settings and real-world applications would be invaluable in broadening our contributions to the development of safer, more reliable, and more efficient collective adaptive systems.
\\[1pt]
To this end, we provide a structured approach to understanding and mitigating emergent behavior, opening new avenues for ensuring the safety and reliability of MAS in increasingly complex and dynamic environments.
Our work has implications for a wide range of applications, from autonomous transportation systems and smart cities to collaborative robotics and intelligent manufacturing. 
It lays the groundwork for more resilient and trustworthy collective adaptive systems that meet the demands of an increasingly interconnected world.

\section*{Acknowledgments}
This work was funded by the Bavarian Ministry for Economic Affairs, Regional Development and Energy as part of a project to support the thematic development of the Institute for Cognitive Systems.

\bibliographystyle{splncsnat}
\bibliography{EMAS}

\end{document}